\documentclass[aps,amsfonts,nofootinbib]{revtex4}

\usepackage{epsf}
\newcommand{\beq}{\begin{equation}}
\newcommand{\eeq}{\end{equation}}
\newcommand{\bea}{\begin{eqnarray}}
\newcommand{\eea}{\end{eqnarray}}
\newcommand{\ce}{{\mathcal{E}}}
\newcommand{\cj}{{\mathcal{J}}}
\newcommand{\cR}{{\mathcal{R}}}

\newcommand{\cQ}{{\mathcal{Q}}}
\newcommand{\cL}{{\mathcal{L}}}

\begin{document}

\title{Separate Universe Approach and the Evolution of
  Nonlinear Superhorizon Cosmological Perturbations}

\author{G.I. Rigopoulos} 
\author{E.P.S. Shellard}

\affiliation{ Department of Applied Mathematics and Theoretical Physics\\
Centre for Mathematical Sciences\\
University of Cambridge\\
Wilberforce Road, Cambridge, CB3 0WA, UK}


\begin{abstract}
\noindent In this letter we review the separate universe approach for
cosmological perturbations and point out
that it is essentially the lowest order approximation to a gradient
expansion. Using this approach, one can study the nonlinear evolution
of inhomogeneous spacetimes and find the conditions under which the
long wavelength curvature perturbation can vary with time. When there
is one degree of freedom or a well-defined equation of state the nonlinear 
long wavelength curvature perturbation remains constant. With
more degrees of freedom it can vary and this variation is determined
by the non-adiabatic pressure perturbation, exactly as in linear
theory. We identify combinations of spatial vectors characterizing the
curvature perturbation which are
invariant under a change of time hypersurfaces.
\end{abstract}

\maketitle

\section{Introduction}    

Until quite recently, a standard tool used in inflationary
calculations was the conservation of the curvature perturbation
$\zeta$ on superhorizon scales. This made inflationary predictions
insensitive to what was going on between horizon crossing and horizon
reentry and thus could lead to very robust predictions. However, 
it was realised \cite{wmll} that in the presence of more degrees of
freedom $\zeta$ can vary on superhorizon scales due to the presence of
a non-adiabatic pressure perturbation. This fact was shown in a very
simple manner in \cite{wmll} using only energy momentum
conservation. The rule for a varying or non-varying $\zeta$ was
derived using linear perturbation theory. The extent to which the rule is
true when nonlinearities are taken into account remained
largely unanswered, although one would expect intuitively that it would
still hold. As far as we are aware, there are three references where nonlinear
conservation of a variable connected to the linear $\zeta$ has been mentioned
\cite{sb,brand,mald} and, in all cases, their results refer to single field inflationary models.              

Dealing with nonlinearities in general relativity is in general a very difficult
issue. Yet there exists an approximation which makes the problem
tractable and which is particularly relevant during inflation. The
quasi-exponential expansion stretches modes to vast superhorizon
scales. Hence, a long wavelength approximation for the study of
inhomogeneous spacetimes smoothed on scales larger than the
horizon becomes particularly relevant. The resulting picture of the
inhomogeneous universe is quite simple. Each point evolves like a
separate homogeneous universe with slightly different values for the
Hubble rate, scalefactor, scalar field  etc. One can find  nonlinear variables
describing the inhomogeneity which do not depend on the choice of time hypersurfaces and which are exactly
gauge-invariant in the perturbation theory sense. They are
combinations of spatial gradients and are immediately
connected with the linear gauge invariant variables $\cR$ and $\zeta$
(see, for example,  \cite{ll} and references therein). In this letter we 
would like to derive such variables
describing the long wavelength curvature perturbation and find their
evolution equations. The latter are essentially the same as those of
linear perturbation theory. In the process we also elaborate
on the separate universe picture which, so far, has only been used heuristically.       

\section{The Long Wavelength Approximation and the Separate Universe Picture}

We start by assuming that we can smooth out all relevant quantities on scales much larger than the
instantaneous comoving hubble radius $\sim(aH)^{-1}$. 
That is, for any quantity Q we will consider the quantity
\label{eq:2}
\beq
\bar{Q}(x)\equiv{}R^{-3}\int{d}^3x'Q(x')W(|x-x'|/R).
\eeq
W is a window function rapidly decreasing for $|x-x'|>R$ and R is a smoothing
scale larger than $(aH)^{-1}$.
We assume that the Einstein equations act on the $\bar{Q}$'s.   
Of course this is not exact. Smoothing
the full equations, i.e convolving the equations with the window
function, is {\it not} obviously equivalent to the original equations acting on
smoothed fields in the case of nonlinear operators. For example, if $Q$ satisfies
\label{eq:3}
\beq
D(Q)=0 
\eeq 
then, in general,
\label{eq:4} 
\beq
\int{d}^3x'D(Q(x'))W(|x-x'|/R)\neq{D}(\bar{Q})
\eeq
if $D$ is a nonlinear operator. Here we assume that $D(\bar{Q})=0$ is an
adequate description of the physics at long wavelengths. Then it is a
reasonable approximation to say that we can drop those terms in the
equations which
contain more than one spatial gradient \cite{sb}. They are
expected to be small compared with time derivatives. This will be the main
approximation and will lead to a picture where different points of the
universe evolve independently \footnote{Particular solutions for long
  wavelength equations as well as an iteration scheme based on a gradient
  expansion has been considered by various authors in the past, see eg.
  \cite{long}. A separate universe
  picture has also been discussed in \cite{separate} in the context of
  perturbing solutions of the homogeneous equations w.r.t the
  constants that appear in them. Here, we are interested not in particular solutions but
  general properties of the nonlinear evolution of the long wavelength
  inhomogeneity codified by variables like (52).}. 

An interesting picture emerges when one also considers the quantum
fluctuations of the various physical fields during an inflationary
era. In such an era the comoving horizon decreases rapidly and the
various modes that are exposed as they exit the horizon can be treated
as classical stochastic fields \cite{star}. Assuming that the equations
hold when acting on the $\bar{Q}$'s and writing the relevant
Einstein equations as first order in time one sees that a time
dependent window function results in extra terms in the
equations. Indeed the time derivative of $\bar{Q}$ now has two terms, 
\beq
{\partial\bar{Q}\over\partial t}=\overline{{\partial Q\over\partial
    t}}+\int{d}^3x'Q(x'){\partial\over\partial t}W(|x-x'|/R)
\eeq       
The integration in (4) is restricted over modes close to horizon exit
so $Q(x')$ can be taken to be the quantum short wavlength
field which has turned classical. Hence all evolution equations get
augmented by a stochastic term. This is what has been used in the past
to derive an effective langevin equation for the evolution of the
inflaton field. In this letter we are concerned with the
evolution of the spacetime after the inhomogeneity has been set up by
quntum fluctuations. We develop a stochastic framework that
takes into account gravitational perturbations and is gauge invariant
in the long wavelength limit in \cite{geri paul}.     

Now, consider the ADM parametrisation of the metric \cite {bardeen,mtw}. One imagines spacetime
to be foliated by spacelike hypersurfaces with a normal vector
\beq
n_{0}=-N,\,\,n_{i}=0,\,\,n^{0}=N^{-1},\,\,n^{i}=N^{-1}N^{i}.
\eeq 
Here, $N$ is called the lapse function and $N^i$ the shift vector. The
latter measures the deviation of the timelike lines which define the
spatial coordinates from the integral curves of the normal to the time
hypersurfaces. The metric
takes the form
\beq
g_{00}=-N^2+\gamma^{ij}N_iN_j\,,\,\,
g_{0i}=-N_i\,,\,\,
g_{ij}=\gamma_{ij}\,,
\eeq
with the inverse
\beq
g^{00}=-N^{-2}\,,\,\,g^{0i}=-N^{-2}N^i\,,\,\,g^{ij}=\gamma^{ij}-N^{-2}N^iN^j\,,
\eeq
where $\gamma_{ij}$ is the metric on the spatial hypersurfaces. The four
functions $N$ and $N^i$ parametrize the gauge freedom of general
relativity and are arbitrary. The choice of gauge in the conventional
language of perturbation theory is essentially the choice of $N$ and
$N^i$ for linearized perturbations. The way the spatial hypersurfaces
are embedded in the 4D geometry is parametrised by the extrinsic
curvature tensor
\beq
K_{ij}=-\frac{1}{2N}\left(N_{i|j}+N_{j|i}+\frac{\partial}{\partial{t}}\gamma_{ij}\right).
\eeq
Here a vertical bar denotes a covariant derivative w.r.t
$\gamma_{ij}$. {\it From now on we will use coordinate systems with 
shift vector $N^i=0$}. For matter described by the energy momentum tensor $T_{\mu\nu}$ we
have the energy density  
\beq
\ce\,{\equiv}\,n^{\mu}n^{\nu}T_{\mu\nu}=N^{-2}T_{00},
\eeq
the momentum density 
\beq
\cj_{i}\equiv-n^{\mu}T_{\mu\,i}=-N^{-1}T_{0i},
\eeq
and the stress tensor
\beq
S_{ij}\equiv{}T_{ij}.
\eeq
A bar above a tensor will denote its traceless part
\beq
\bar{K}_{ij}=K_{ij} - \frac{1}{3}K\gamma_{ij}\, ,\qquad K=K^i_{\,i}=\gamma^{ij}K_{ij}.
\eeq  
We can now write the Einstein equations for the above
variables, dropping all terms which explicitely contain second order
spatial derivatives and setting $N^i$=0. The $00$ and $0i$ Einstein
equations are 
\bea
\bar{K}_{ij}\bar{K}^{ij}&-&\frac{2}{3}K^2+\frac{16\pi}{m_{\rm pl}^2}\ce
=0\,,\\
\bar{K}^{j}_{\,i|j}&-&\frac{2}{3}K_{|i}-\frac{8\pi}{m_{\rm pl}^2}\cj_{i}=0\,.
\eea
The first will turn out to be equivalent to the Friedmann equation of
the FRW cosmology and the second is usually refered to as the momentum
constraint. The dynamical equations are
\bea
\frac{d{K}}{d{t}}&=&N\frac{3}{2}\bar{K}_{ij}\bar{K}^{ij}+\frac{12\pi}{m_{\rm pl}^2}N\left(\ce+\frac{1}{3}S\right)\,,\\
\frac{d{\bar{K}^{i}_{\,j}}}{d{t}}&=&N\left(\bar{K\bar{K}^{i}_{\,j}}-\frac{8\pi}{m_{\rm pl}^2}\bar{S}^{i}_{\,k}\right)\,,
\eea
where $\bar{S}_{ij} = {S}_{ij} - {1\over3} S\gamma_{ij}$ with pressure
$S = S^{i}_{\,i} = \gamma^{ij}S_{ij}$ and $\cR^i_j$ is the Ricci tensor of
the spatial hypersurfaces.  
Matter will obey the continuity equations $T^{\mu\nu}{}_{;\nu}=0$
which take the form
\bea 
\frac{d\ce}{d{t}}&=&NK(\ce+\frac{1}{3}S)+N\bar{K}^{ij}\bar{S}_{ij}+N^{-1}(N^2\cj^{i})_{|i}\,,\\
\frac{d\cj_i}{d{}t}&=&NK\cj_i-(\ce\delta^j_i+S^j_i)N_{|j}-NS^j_{i|j}.
\eea
In the case of matter composed of several scalar fields $\phi^A$, as
is relevant in scalar field driven inflation, the
energy momentum tensor is 
\beq
T_{\mu\nu}=G_{AB}\partial_{\mu}\phi^A\partial_{\nu}\phi^B-g_{\mu\nu}\left(\frac{1}{2}G_{AB}\partial^{\lambda}\phi^A\partial_{\lambda}\phi^B+V\right),
\eeq
so that
\bea
\ce&\simeq&\frac{1}{2N^2}G_{AB}\dot\phi^A\dot\phi^B+V(\phi)\,,\\
\cj_i&=&-\frac{1}{N}G_{AB}\dot\phi^A\partial_i\phi^B\,,\\
S_{ij}&\simeq&\gamma_{ij}\left(\frac{1}{2N^2}G_{AB}\dot\phi^A\dot\phi^B-V\right),
\eea
where we have dropped second order spatial gradients. So in the case
of a collection of scalar fields in the long wavelength approximation
$(N^2\cj_i)_{|j}\simeq0$ and $\bar{S}_{ij}\simeq0$, up to second order
spatial gradients. Actually, for a general perfect
fluid
\beq 
T_{\mu\nu}=(\ce+p)u_{\mu}u_{\nu}+pg_{\mu\nu}.
\eeq
So, in the absence of pure vector perturbations on super-Hubble
scales, $u_i=\partial_if$ and the long wavelength approximation gives
$S^i{}_j\simeq{}p\delta^i{}_j$, $\bar{S}^i{}_j\simeq0$ and
$\cj_{i|j}\simeq0$. 
The set of relevant evolution equations then become
\bea
\frac{d{K}}{d{t}}&=&N\frac{3}{2}\bar{K}_{ij}\bar{K}^{ij}+\frac{12\pi}{m_{\rm pl}^2}N\left(\ce+\frac{1}{3}S\right)\,,\\
\frac{d{\bar{K}^{i}_{\,j}}}{d{t}}&=&NK\bar{K}^{i}_{\,j}\,,\\
\frac{d\ce}{dt}&=&NK\left(\ce+\frac{1}{3}S\right)\,,\\
\frac{d\cj_i}{d{}t}&=&NK\cj_i-(\ce\delta^j_i+S^j_i)N_{|j}-NS^j_{i|j}\,,
\eea
which, along with the constraints
\bea
\bar{K}_{ij}\bar{K}^{ij}&-&\frac{2}{3}K^2+\frac{16\pi}{m_{\rm pl}^2}\ce
=0\,,\\
\bar{K}^{j}_{\,i|j}&-&\frac{2}{3}K_{|i}-\frac{8\pi}{m_{\rm pl}^2}\cj_{i}=0
\eea
form the basis of this approach. 

It will now be convenient to write the spatial metric as \cite{bardeen}
\beq
\gamma_{ij}=\gamma^{1/3}\tilde\gamma_{ij}
\eeq
where $\gamma$ is the determinant of the spatial metric, taking 
$Det(\tilde\gamma_{ij})\equiv1$. Now since $Tr(\dot{\tilde\gamma}_{ij})=0$, 
we have
\beq
\bar{K}_{ij}=-\frac{1}{2N}\gamma^{1/3}\dot{\tilde\gamma}_{ij}
\eeq
and 
\label{eq:hubble}
\beq
K=-\frac{1}{2N}{\dot\gamma\over\gamma}.
\eeq
The determinant $\gamma(\mathbf{x},t)$ contains what in linear theory
is usually called a scalar
perturbation mode of the metric. $\tilde{\gamma}_{ij}(\mathbf{x},t)$
contains another scalar, the vector and
tensor perturbations. From (32) we see that
$K(\mathbf{x},t)$ can be interpreted
as a locally-defined Hubble parameter since $\gamma^{1/3}(\mathbf{x},t)\equiv{}a(\mathbf{x},t)$ is a
locally defined scalefactor. It can be explicitly shown
that the system (24-29), a truncated form of Einstein's equations,
forms a consistent set of equations, by which we mean that the evolution
preserves the constraints. So an initial inhomogeneous configuration which respects the constraint (29), can be evolved
with the evolution equations and the constraint always will be satisfied. Now observe that the only equation where spatial
derivatives appear is the constraint (29) (and eqn (27) but this turns
out to be irrelevant). Hence, after initial data
which respect it have
been specified, each point can be evolved individually. The relevant
equations are essentially those of the homogeneous FRW cosmology,
valid locally, modulo the $\bar{K}^i{}_j$ terms. This is exactly what 
has been termed in the past as the `separate
universe approach' (see, for example, \cite{wmll}). It was explicitly
stated in \cite{sb} in the context of general relativistic
Hamilton-Jacobi theory. We see that such an approach is
equivalent to the first order of a spatial gradient expansion.      

Although we can solve the equations numerically retaining the
$\bar{K}^i{}_j$ terms, we can readily see that they are not expected
to be important dynamically \cite{sb}. Using (32) we can solve
(25) to get
\beq
\bar{K}^i{}_j=C^i{}_j(\mathbf{x})\gamma^{-1/2}.    
\eeq
So, in the absence of sources, in an expanding universe (particularly for
quasi-exponential expansion), the traceless
part of the extrinsic curvature $\bar{K}^i{}_j$ decays
rapidly. In most cases, therefore, it will be
safe to ignore the $\bar{K}$ terms. Of course, in the presence of
quantum fluctuations $\bar{K}$ is sourced mostly by the fluctuating
$\gamma$ (see (31)).
Ignoring $\bar{K}$, equations (24-29) become
\bea
\frac{dK}{dt}&=&\frac{12\pi}{m_{\rm pl}^2}N\left(G_{AB}\Pi^A\Pi^B\right)\\
\frac{d}{dt}\Pi^A&=&NK\Pi^A-N\Gamma^A_{BC}\Pi^B\Pi^C-NG^{AB}
{\partial V\over\partial\phi^B}\\
K^2&=&\frac{24\pi}{m_{\rm pl}^2}\left(\frac{1}{2}G_{AB}\Pi^A\Pi^B+V\right)\\
\partial_iK&=&\frac{12\pi}{m_{\rm pl}^2}G_{AB}\Pi^A\partial_i\phi^B,
\eea
where 
\beq
\frac{\dot\phi^A}{N}=\Pi^A,
\eeq 
and $\Gamma^A_{BC}$ is the Connection formed from $G_{AB}$. 
Eqns (34-37) are exactly the same as those of a homogeneous
cosmology apart from the momentum constraint (37). Hence, the
long wavelength universe looks like a collection of homogeneous universes, each
evolving according to the equations of FRW cosmology. The only
information about inhomogeneity is contained in the momentum
constraint (37) \cite{sb}.    

In single field homogeneous cosmology, it is very convenient to
parametrize the Hubble rate $H$ in terms of the value of the scalar
field leading to what is called the Hamilton Jacobi formulation. A
similar thing can be done here. Observe that (37)
is satisfied if $K=K(\phi^A)$ and 
\beq
\Pi^A=\frac{m_{\rm pl}^2}{12\pi}G^{AB}\frac{\partial{K}}{\partial\phi^B}.
\eeq
Substituting this into the Friedmann equation (36) gives the Hamilton
Jacobi equation for a system of scalar fields \cite{sb}
\beq
 K^2-\frac{m_{\rm pl}^2}{12\pi}G^{AB}\frac{\partial{K}}{\partial\phi^A}\frac{\partial{K}}{\partial\phi^B}=\frac{24\pi}{m_{\rm pl}^2}V.
\eeq
So, $K$ is a function of the scalar fields only. Note that this is
consistent with equation (34) which can now be written
\beq
\dot{K}=\frac{\partial{K}}{\partial\phi^C}\dot{\phi^C}.   
\eeq
Hence, given appropriate boundary conditions, a solution to equation (40) determines
the state of the system completely in terms of its position in the
configuration space of the scalar fields. This result now holds for an
inhomogeneous universe smoothed on superhorizon scales.

Solving (40) analytically is not straightforward for a
general potential. Explicit solutions for the case of an exponential
potential have been given in \cite{sb}. During inflation, it is usually
a very good approximation to consider a slow-roll behaviour where the
derivative terms can be neglected. Then (40) becomes the usual slow-roll 
relation 
\beq
K^2=\frac{24\pi}{m_{\rm pl}^2}V.
\eeq
In this case the momentum constraint is equivalent to (36). Another case
where an approximate solution can be found is the study of deviations
of  the scalar field about some stable point which are
small compared to $m_{\rm pl}$ \cite{kin}. In any case, a solution to
equation (40) describes the long wavelength system completely. An
interesting point noted in \cite{sb}, is the 
attractor property of solutions of (40). The latter contain a number
of arbitrary parameters $C^A$, reflecting the
freedom to choose initial conditions of the field momenta (to specify
the motion one must assign not only an initial value for $\phi$ but
also $\dot\phi$ at each point). Neglecting
the $\bar{K}$ terms means that these parameters are spatially
independent. Of course, when assigning initial conditions the momenta
need not be strictly determined by the field values. So, in principle
these constant parameters should be spatially dependent too. So
suppose one considers a function $K(\phi^A,C)$, then
\bea
2K\frac{\partial{K}}{\partial{C}}&=&\frac{m_{\rm pl}^2}{6\pi}G^{AB}\frac{\partial{K}}{\partial\phi^A}\frac{\partial}{\partial\phi^B}\frac{\partial{K}}{\partial{C}}\nonumber\\
\Rightarrow{}K&=&\frac{m_{\rm pl}^2}{12\pi}G^{AB}\frac{\partial{K}}{\partial\phi^A}\frac{\partial}{\partial\phi^B}\ln{\left|\frac{\partial{K}}{\partial{C}}\right|}\nonumber\\
&=&\frac{1}{N}\frac{\partial}{\partial{t}}\ln{\left|\frac{\partial{K}}{\partial{C}}\right|}
\eea
where we have used (38) and (39). From (43) and (32) we see that
\beq
\left|\frac{\partial{K}}{\partial{C}}\right|\propto{}a^{-3}.
\eeq
Thus the freedom to choose the momentum independently at each point,
reflected in the freedom to choose constants in the solution of (40),
becomes irrelevant very quickly\footnote{This is intimately connected
to the fact that perturbation modes freeze when they exit the horizon}.   

From now on we will also be using the more familiar notation 
\bea
K(t,\mathbf{x})&=&-3H(t,\mathbf{x})\,,\nonumber\\
\gamma(t,\mathbf{x})&=&a^6(t,\mathbf{x}),
\eea
where $a$ and $H$ are the locally defined values of the scale factor and
the 
Hubble rate.

\section{The Long Wavelength Curvature Peturbation}

In linear perturbation theory, a common variable used to characterize
the perturbations is $\delta\zeta$ defined as
\beq 
\delta\zeta=\psi-H\frac{\delta\ce}{\dot\ce}=\psi-\frac{H}{\dot{H}}\left(\dot\psi+H\psi\right)=\psi-H\frac{\delta{}H}{\dot{H}}
\eeq
where $\psi$ is the scalar perturbation of the spatial part of the
metric, $\psi={\delta{}a(t,\mathbf{x})}/{a(t)}$. The quantity 
$\delta\zeta$ is gauge invariant and corresponds to the curvature
perturbation in a time slicing that sets $\delta\ce=0$. In the case
of single field inflation, where there is one degree of freedom, on
long wavelengths,
$\delta\dot\zeta=0$. When there are many scalar fields present \cite{wmll}
\beq
\delta\dot\zeta=-\frac{H}{(\ce+p)}\delta{}p_{nad},
\eeq
where $\delta{}p_{nad}$ is the non-adiabatic pressure perturbation. In
general, when there is a well defined equation of state $p=p(\ce)$,
$\delta{}p_{nad}$=0. We will now derive a nonlinear generalisation of $\delta\zeta$
and eqn (47) for the nonlinear long wavelength case. 

Consider a generic time slicing and some initial time $t_I$. We can
integrate the local metric determinant up to a surface of constant
energy density using (32)   
\beq
\ln{a(t_I,\mathbf{x})}=\ln{a(t,\mathbf{x})}-\int\limits^{T}_{t_I}N(t',\mathbf{x})H(t',\mathbf{x})dt',
\eeq
where $T\left(\ce(\mathbf{x})\right)$ is the value of $t$ at the point
$\mathbf{x}$ for $\ce(\mathbf{x})=const$. Taking the spatial derivative and using (32) once more we see that
\bea
\partial_i\left[\ln{a(t_I,\mathbf{x})}\right]&=&\left(\partial_i-\frac{\partial{}T}{\partial{}x^i}\partial_t\right)\ln{a(t,\mathbf{x})}-\int\limits^{T}_{t_I}\partial_i\left[N(t',\mathbf{x})H(t',\mathbf{x})\right]dt'\,\nonumber\\
&=&\left(\partial_i-\frac{\partial{}T}{\partial{}x^i}\partial_t\right)\ln{\gamma(t,\mathbf{x})}-\int\limits^{T}_{t_I}\partial_i\partial_t\left[\ln{a(t',\mathbf{x})}\right]\,dt'
\eea 
which can be written 
\beq
\partial_i[\ln{a(T,\mathbf{x})}]=\partial_i[\ln{a(t,\mathbf{x})}]-\frac{\partial{}T}{\partial{}x^i}N(t,\mathbf{x})H(t,\mathbf{x}),
\eeq 
or 
\beq
X_i^{(\ce)}=X_i-\frac{NH}{d\ce/dT}\partial_i\ce,
\eeq
where we have defined $X_i\equiv\partial_i\ln{a}$ and $T=T(\ce)$. Note that the
gradient of the l.h.s of (50) is evaluated along the hypersurface
$\ce=const$. We see that with any choice of $N$, the particular
combination of gradients appearing on the r.h.s of (51) always equals
the gradient of $\ln{a}$ on a surface where $\ce=const$. So we are led to examine the variable
\beq
\zeta_i=X_i-\frac{NH}{\dot\ce}\partial_i\ce,
\eeq 
as a coordinate independent measure of the nonlinear curvature
perturbation. Let us first see how this
variable evolves with time. Taking the time derivative of (52) and using (26)
we arrive at
\beq    
\dot\zeta_i=\frac{1}{3}\frac{1}{\left(\ce+\frac{1}{3}S\right)^2}\left(\dot\ce\partial_i{}S-\dot{S}\partial_i\ce\right),
\eeq
or, by using (26) once more
\beq
\dot\zeta_i=-\frac{1}{3}\frac{NH}{\left(\ce+\frac{1}{3}S\right)}\left(\partial_iS-\frac{\dot{S}}{\dot\ce}\partial_i\ce\right).
\eeq
This should be compared with (47). It is easy to see that if there is
a well defined equation of state $S=S(\ce)$, then the r.h.s of (53) is zero. Therefore, the long
wavelength curvature perturbation, described by the vector $\zeta_i$, can evolve on superhorizon scales if and only
if there exists a  non-adiabatic pressure perturbation. This was proved in \cite{wmll} for the case of linear perturbation
theory by making use only of energy
conservation. The same applied here. Hence equation (28) shows that
this result extends to the nonlinear case if we use the vector
$\zeta_i$ to describe the perturbation. 

The connection with the usual linearized
gauge invariant variable $\zeta$ is obvious from (52). Writing $X=\ln(\gamma)$, $\ce=\ce(t)+\delta\ce$,
$\gamma=\gamma(t)+\delta\gamma$ and dropping terms quadratic in
the perturbations, we see that
\bea
\zeta_i&\simeq&\partial_i\delta\zeta\,,\nonumber\\
\delta\zeta&\simeq&\frac{\delta{a}}{a}-\left(\frac{NH}{\dot\ce}\right)(t,\mathbf{x}_0)\,\delta\ce+C(t),
\eea
with C(t) the background number of e-folds.

The quantity $\zeta_i$ was defined with respect to the energy
density. Also the way we arrived at it was at best heuristic. We will now show that for any inhomogeneous spacetime scalar
$\phi(t,\mathbf{x})$, the variable
\beq
\cR_i=X_i-\frac{NH}{\dot\phi}\partial_i\phi    
\eeq
is gauge invariant in the long wavelength approximation under mild
conditions regarding the allowed coordinate transformations. Consider the
transformation
\beq
(t,\mathbf{x})\rightarrow\left (T(t,\mathbf{x}),\mathbf{X}(t,\mathbf{x})\right).
\eeq
Then, it can be shown that \cite{sb}
\beq
X^j=f^j(x^i)+\int\frac{T^{,j}}{T^{,0}T_{,0}}dT,
\eeq
where $f^j$ is an arbitrary function independent of time. One can
choose it for convenience to be $\delta^i_jx^j$ \cite{sb} but one can also
attach a physical meaning to such a choice as can be seen in figure 1. 
Related considerations in matching perturbed and homogeneous spacetimes are
discussed in \cite{AmeryShell}.  We
note that the transformation (58) holds between gauges with the shift $N_i=0$. Assuming that the new time coordinate $T$ is non-singular, that is, that the
new time hypersurfaces do not get too `wrinkly', we can discard the
gradient of the integral as second order in spatial gradients. We then have  
\beq
\frac{\partial{}X^j}{\partial{}x^i}\simeq\delta^j{}_i.
\eeq
Consider now a gauge with coordinates $x^{\mu}$ where
$\partial_i\phi=0$ and a transformation to a different gauge with
coordinates $X^{\tilde\mu}$. Then
\bea
\gamma_{kl}&=&\frac{\partial{}X^{\tilde\mu}}{\partial{}x^k}\frac{\partial{}X^{\tilde\nu}}{\partial{}x^l}g_{\tilde\mu\tilde\nu}\nonumber\\
&\simeq&-T_{,k}T_{,l}N^2_T+\delta^{\tilde{k}}{}_k\delta^{\tilde{l}}{}_l\gamma_{\tilde{k}\tilde{l}}\nonumber\\
&\simeq&\delta^{\tilde{k}}{}_k\delta^{\tilde{l}}{}_l\gamma_{\tilde{k}\tilde{l}}\,,\eea
and so
\beq
\gamma(t,\mathbf{x})\simeq\tilde{\gamma}(T,\mathbf{X}).
\eeq
Taking the spatial derivative w.r.t $x^i$ we get
\bea
\frac{\partial}{\partial{}x^i}\ln\gamma&=&\frac{\partial{T}}{\partial{}x^i}\partial_T\ln\tilde\gamma+\frac{\partial{}X^j}{\partial{}x^i}\frac{\partial}{\partial{}X^j}\ln\tilde\gamma\nonumber\\
&\simeq&\partial_iT\partial_T\ln\tilde\gamma+\frac{\partial}{\partial{}X^i}\ln\tilde\gamma.
\eea

\begin{figure}[t]

\centerline{\epsfxsize=6cm\epsfbox{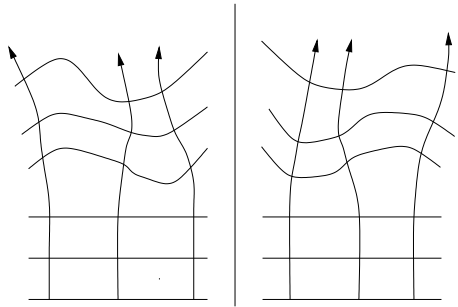}}

\caption{Two different slicings of an inhomogeneous spacetime based
  on two different time variables, $T$ and $\tilde{T}$, have different
  normal curves defining the spatial coordinates. A particular patch
  of spacetime during inflation starts its life inside the horizon
  where it can be considered as homogeneous. Classical perturbations
  are generated when modes cross the horizon and freeze in. While
  subhorizon and homogeneous, there is a preferred
  time slicing in the patch to which all slicings and spatial
  coordinate choices should match. This
  fixes $f^j(x^i)=\delta^j_ix^i$. This choice separates changes in the
  time slicing -- which we want to consider as a gauge choice -- from possible coordinate
  transformations of the homogeneous spacetime.}

\end{figure}

For the scalar $\phi(t,\mathbf{x})$ we have $\partial_i\phi=0$, so in
the new coordinates
\beq
\partial_iT\partial_T\tilde\phi(T,\mathbf{X})+\frac{\partial}{\partial{}X^i}\tilde\phi(T,\mathbf{X})=0.
\eeq
Therefore 
\beq
\partial_i\ln\gamma=\frac{\partial}{\partial{}X^i}\ln\tilde\gamma-\frac{\partial_T\ln\tilde\gamma}{\partial_T\phi}\frac{\partial}{\partial{}X^i}\phi\,.
\eeq
Now, since the transformation we used was arbitrary (up
to considerations for the
smooth behaviour of the new time $T$), and noting that 
\beq
H=-\frac{1}{6N_T}\partial_T\ln\tilde\gamma
\eeq
and $X_i = \partial_i \ln a$, we conclude that the variable 
\beq
\mathcal{R}_i=X_i-\frac{NH}{\dot\phi}\partial_i\phi,   
\eeq
equals the gradient of the spatial metric determinant in a gauge where
the scalar $\phi$ is homogeneous;  it is, therefore, a nonlinear gauge
invariant measure of the perturbations. When the scalar is taken to be
the energy density $\ce$ we recover $\zeta_i$ of eqn (52). 

We would like to remark here that the use of spatial gradients to
describe perturbations in a gauge invariant manner was first advocated in
\cite{ellis}. There, the use of the word `gauge' refers to a choice of
a correspondence between a fictitious background homogeneous spacetime
and the real perturbed universe. So a gauge invariant quantity is one that does
not change when this correspondence is altered, which is achieved
when the quantity either vanishes or is a trivial tensor in the
background \cite{j.m.s,bruni}. In this sense,
any gradient is gauge invariant since it must vanish on the homogeneous 
background. Here, however, by a choice of gauge we mean a choice of the 
lapse function $N$ and the
shift vector $N_i$ \cite{ll},  so we look for quantities such as (66) that are
invariant under a change of time slicing as well.

\medskip
\noindent {\it Examples}
\smallskip

When the scalar $\phi$ in (56) is taken to be the inflaton it can be easily verifield
that $\cR_i$ is also conserved. Using eqns (34--37) one can show
that 
\beq
\dot\cR_i=0\,.
\eeq
When there is more than one scalar field present, however, the situation is
different. It is well known in linear perturbation theory \cite{wmll,
gordon, geri} that perturbations in multi-scalar field models can source the curvature
perturbation $\delta\zeta$ on long wavelengths, making the time evolution of the
latter possible. Let us see how this effect is very easily seen in the
formalism developed in the previous section without the need to resort
to linear perturbation theory. Consider
hypersurfaces with $\partial_i\ce=0$. From (28) with the $\bar{K}$
terms omitted we see that
\beq
\ce_{,i}=0\Rightarrow{}H_{,i}=0\Rightarrow{}\cj_i=0\Rightarrow{}G_{AB}{\Pi^A}\partial_i\phi^B=0.
\eeq
The fact that $\ce_{,i}=0$ also means that
\beq
S_{,i}=2G_{AB}{\Pi^A}\partial_i{\Pi^B}.
\eeq
So, if there is a single field,
\beq
\ce_{,i}=0~\Rightarrow{}~H_{,i}=0~\Rightarrow{}~\partial_i\phi=0
~\Rightarrow{}~S_{,i}=0\,,
\eeq
since $\partial_i{\Pi}=0$. Hence $\zeta_i$ remains constant. On the other hand, 
for many fields, we have
\beq
\ce_{,i}=0~\Rightarrow{}~H_{,i}=0~\Rightarrow{}~G_{AB}{\Pi^A}\partial_i\phi^B=0.
\eeq
So, generally in this case, we may have $\partial_i\phi^A\neq0$. Therefore, from
(69) $\partial_iS\neq0$ in general and $\zeta_i$ can evolve on superhorizon
scales, $\dot\zeta_i\neq0$.

In general, for a system of scalar fields governed by the Lagrangian 
\beq
\cL=\frac{1}{2}G_{AB}\partial^\mu\phi^A\partial_\mu\phi^B-V,
\eeq
the variation of $\zeta_i$ from (54) is given by
\beq
\dot\zeta_i=V_A\left[\frac{d}{dt}\left(\frac{1}{G_{BC}\Pi^B\Pi^C}\right)\partial_i\phi^A-\dot\phi^A\partial_i\left(\frac{1}{G_{BC}\Pi^B\Pi^C}\right)\right].
\eeq
In the multi field case it is convenient to consider the variables
\beq
\cQ_i^A=\Pi^AX_i-H\partial_i\phi^A,
\eeq
which are also invariant under changes of time hypersurfaces in the
long wavelength limit. They are not conserved even in the case of a single field. In
particular
\beq
\dot\cQ_i^A=\dot\Pi^A\cR_i^A+\Pi^A\dot\cR_i^A,
\eeq
where no summation is implied. It should be noted here that it is possible to define conserved
variables for each field if and only if the potential is of the form
$\sum{}V\left(\phi^A\right)$ or if one field dominates completely over
the other. In terms of the $\cQ^A_i$'s the curvature
perturbation is given by
\beq
\zeta_i=\frac{G_{AB}\Pi^A\cQ^B_i}{G_{CD}\Pi^C\Pi^D}. 
\eeq
Consider now the following simple two-field example which can be considered a toy
model for hybrid inflation. Initially the first field $\sigma$ is stuck
slightly displaced from the bottom of a valley while the other field $\phi$ evolves and drives
inflation. In this case, we have essentially a single field model and
$\zeta_i$ is conserved. With $\Pi^\sigma\simeq0$, we have
\beq
\zeta^{(I)}_i=X_i-\frac{H}{\Pi^\phi}\left(\partial_i\phi+\frac{G_{\phi\sigma}}{G_{\phi\phi}}\partial_i\sigma\right)\,.
\eeq   
When the $\phi$ field reaches a critical value, the trajectory turns
abruptly in the direction of variations in $\sigma$ which can now dominate until
inflation ends. For the period of rolling in the $\sigma$-direction we have an extra
contribution to $\zeta_i$
 \beq
\zeta^{(II)}_i=X_i-\frac{H}{\Pi^\sigma}\left(\partial_i\sigma+\frac{G_{\phi\sigma}}{G_{\sigma\sigma}}\partial_i\phi\right).
\eeq
Note that for a non-diagonal metric $G_{AB}$ the isocurvature $\sigma$-field can
contribute to the curvature perturbation even for a straight
trajectory \cite{br,geri}, in contrast to what happens when
$G_{AB}=\delta_{AB}$ \cite{gordon}. Also, if during the rolling
along the $\sigma$ direction we have a
second period of inflation, the $\sigma$ field can contribute to the curvature
perturbation since
\beq
\zeta_i=\zeta^{(I)}_i+\zeta^{(II)}_i.
\eeq
In the case with $G_{AB}=\delta_{AB}$, we can easily find the following evolution
equations for the $\cR_i$'s in the case of two fields
\beq
\dot\cR^\phi_i=\frac{4\pi}{m_{\rm pl}^2}N\Pi_\sigma\left(\partial_i\sigma-\frac{\Pi_\sigma}{\Pi_\phi}\partial_i\phi\right)
\eeq
and
\beq
\dot\cR^\sigma_i=\frac{4\pi}{m_{\rm pl}^2}N\Pi_\phi\left(\partial_i\phi-\frac{\Pi_\phi}{\Pi_\sigma}\partial_i\sigma\right)\,.
\eeq

For many general fluids one can write the curvature
perturbation as a weighted sum over components \cite{LythWand}
\beq
\zeta_i=\frac{\sum\dot\ce_A\zeta^A_i}{\sum\dot\ce_A}.  
\eeq
If the components are non interacting, each will obey an evolution
equation like (26), so each $\zeta^A_i$ is separately conserved. Hence
one can see that the time evolution of the total curvature
perturbation is determined by the ratio of the energy densities of the
various components. For example, starting with $\zeta_i^A=0$ we get
\beq
\zeta_i=\frac{\dot\ce_B}{\dot\ce_A+\dot\ce_B}\zeta^B_i 
\eeq

This evolution of the long wavelength curvature perturbation in
multifield models is one of the possible mechanisms that could potentially produce significant non
gaussianity. Perturbations from the $\sigma$ direction can be significantly
non-gaussian and now there is the possibility that these fluctuations
can contribute to the curvature perturbation at later times
\cite{linde mukhanov, bu, luw}. This can be seen for example from
equations (80) and (82). Other mechanisms relating non-gaussianity with a
second scalar field have been proposed \cite{zal}. We will return
on the issue with more details in \cite{geri paul}. 

One can also easily see that there is a particular form of a multi-scalar field
potential which will conserve $\zeta_i$. With $H_{,i}=0$  we have
$G_{AB}{\Pi^A}\partial_i\phi^B=0$, so 
to get $S_{,i}=0$ we need (see (69) and (39))
\beq
G_{AB}{\Pi^A}\partial_i{\Pi^B}=\frac{m_{\rm pl}^2}{12\pi}G_{AB}{\Pi^A}\partial_i\left(G^{BC}\frac{\partial{}H}{\partial\phi^C}\right)=0\,.
\eeq
Thus, if 
\beq
G^{AC}\frac{\partial{}H}{\partial\phi^C}=\frac{\lambda}{m_{\rm pl}}\phi^A\,\Rightarrow\,\frac{\partial{}H}{\partial\phi^A}=\frac{\lambda}{m_{\rm pl}}G_{AB}\phi^B,
\eeq
with $\lambda$ some dimensionless number, then $S_{,i}=0$. From (40)
\beq
H^2=\frac{m_{\rm pl}^2}{12\pi}G^{AB}\frac{\partial{}H}{\partial\phi^A}\frac{\partial{}H}{\partial\phi^B}+\frac{8\pi}{3m_{\rm pl}^2}V(\phi).
\eeq
So, with 
\beq
H=\frac{\lambda}{m_{\rm pl}}\frac{1}{2}G_{AB}\phi^A\phi^B+\mu{}m_{\rm pl},
\eeq
we get
\beq
V(\phi)=-\frac{1}{2}M^2\phi^A\phi_A+M^2\frac{\pi}{2m_{\rm pl}^2}\left(\frac{1}{2}\phi^A\phi_A+qm_{\rm pl}^2\right)^2,
\eeq
with $q$ another dimensionless parameter and $M^2$ can have either sign. So, for a general potential containing both quartic and quadratic
terms with the various constants related as in (76), the long
wavelength curvature perturbation does not evolve in time. One should
keep in mind that the above results only apply when anisotropic stresses are
negligible on superhorizon scales and therefore one can safely neglect
the $\bar{K}^i_j$ terms.

We emphasise again that for
the derivation of all of the above no linear approximation has been
made. The equations are exact and hold at every point of the
inhomogeneous spacetime in the long wavelength limit. For more uses of this formalism we refer the reader to \cite{geri paul}.

\section{Summary}
We have shown that particular combinations of spatial gradients are
good variables to describe the curvature perturbation in a long
wavelength limit. We also elaborated on the intuitive picture of the
`separate universe approach' where, given initial data which respect
the energy and momentum constraints, one can view the long wavelength
universe as a collection of universes each evolving
independently. Such a picture is a simple way to understand 
nonlinear evolution on large scales and, in particular, to determine the 
conditions under which the long wavelength curvature perturbation can vary. 
We believe this is a remarkably straightforward and physically intuitive
approach which should find much wider application.

\bigskip
\noindent{\it Note added:}  While this paper was undergoing final revision, 
independent work by Lyth and Wands appeared on the archive \cite{LythWand}.  
This addresses some related issues on $\zeta$-conservation using linear 
perturbation methods.

\end{document}